\documentstyle[preprint,aps,epsfig]{revtex}
\begin{document}
%=========================================================
\title{
Analysis of
$\Lambda_b \rightarrow \Lambda_c$ weak decays\\
in Heavy Quark Effective Theory
}
%=========================================================
\author{
Jong-Phil Lee\footnote{jplee@phya.snu.ac.kr},
Chun Liu\footnote{liuc@ctp.snu.ac.kr} 
and 
H. S. Song\footnote{hssong@physs.snu.ac.kr}}
%=========================================================
\address{
Department of Physics and Center for Theoretical Physcis,\\
Seoul National University, Seoul, 151-742, Korea }
%=========================================================
\maketitle
%\vspace*{4cm}
\begin{abstract}
The $\Lambda_b\to\Lambda_c$ semileptonic decay is analyzed in the
framework of heavy quark effective theory to the order of
$1/m_c$ and $1/m_b$. 
The QCD sum rule and large $N_c$ predictions to the decay
form factors are applied.
It argues that the subleading baryonic Isgur-Wise function in the
large $N_c$ limit vanishes.
The decay rates, distributions and asymmetry parameters are
calculated numerically. 
Some of the nonleptonic decay modes are discussed in the end.
\end{abstract}
\pacs{13.30.Ce, 14.20.Mr}
%%%%%%%%%%%%%%%%%%%%%%%%%%%%%%%%%%%%%%%%%%%%%%%%%%%%%%%%
%=========================
\section{Introduction}
%=========================
The weak decays of heavy baryons provide testing ground for the Standard
Model.  They reveal some important features of the physics of heavy quarks.
From the study of the heavy quark physics, some important parameters of the 
Standard Model, for instance, the Cabbibo-Kobayashi-Maskawa (CKM) matrix 
element $V_{cb}$ can be extracted by comparing experiments with theoretical 
calculations from the decay mode 
$\Lambda_b \rightarrow \Lambda_c l\bar{\nu}$.
\indent
The main difficulties in the Standard Model calculations,
however, are due to the
poor understanding of the nonperturbative aspects of the strong interactions 
(QCD).\par
For the heavy hadrons containing a single heavy quark, 
an effective theory of QCD 
based on the heavy quark symmetry in the heavy quark limit \cite{IsgurL}, 
the so-called heavy quark effective theory (HQET), 
has been proposed \cite{GeorgiL}.  
The classification of the weak decay form factors of heavy baryons has been
simplified greatly in HQET \cite{WiseL}.  
At the leading order of heavy quark expansion, 
only one universal form factor, the Isgur-Wise function, is required to
describe the $\Lambda_b\to\Lambda_c$ semileptonic decay.
To the order of $1/m_Q$ \cite{LukeL}, one more universal function 
and one mass parameter are introduced \cite{GrinsteinL}.
However, the heavy quark symmetry itself has no power 
to give information about the
details of the universal form factors and the mass parameter.
For a complete analysis to the 
heavy baryons, we still need to employ some other nonperturbative methods.
Interesting results about the heavy baryon weak decay form factors have been 
obtained by various nonperturbative methods.  
They are QCD sum rules \cite{Grozin,Dai},
large $N_c$ limit \cite{ManoharL}, lattice simulation \cite{UKQCD}, 
dispersion relation and analyticity \cite{Boyd}, 
and quark models \cite{IvanovL}.  
\par
In this paper, we apply the results of QCD sum rules and large $N_c$ limit
to analyze in detail the weak decays of $\Lambda_b \rightarrow \Lambda_c$
to the order of $1/m_c$ and $1/m_b$.
The analysis is useful to experiments in the near future. 
In Sec.2, $\Lambda_b\to\Lambda_c$ semileptonic decay form factors 
are discussed.
While there is no large $N_c$ calculation for the universal form factor
appeared in $1/m_Q$ corrections,
we argue that it is zero in the large $N_c$ limit.
In Sec.3, the numerical results for the decay rates, distributions 
and various angular asymmetry parameters are calculated.
In Sec.4, several non-leptonic decay modes of $\Lambda_b$ are discussed.
We summarize the results in Sec.5.
%%%%%%%%%%%%%%%%%%%%%%%%%%%%%%%%%%%%%%%%%%%%%%%%%%%%%%%%%%%%%%%%%%%%%%%%%%%%%%%%
\section{Form Factors}
%%%%%%%%%%%%%%%%%%%%%%%%%%%%%%%%%%%%%%%%%%%%%%%%%%%%%%%%%%%%%%%%%%%%%%%%%%%%%%%%
The hadronic matrix element of the weak current 
appeared in the effective Hamiltonian for $\Lambda_b \rightarrow \Lambda_c$
is parameterized generally by six form factors $F_i$ and $G_i$ 
($i=1, 2, 3$),
\begin{equation}
\begin{array}{lll}
\langle\Lambda_c(v')|\bar{c}\gamma^{\mu}(1-\gamma^5)b|\Lambda_b(v)\rangle&=&
\bar{u}_{\Lambda_c}(v')(F_1\gamma^{\mu}+F_2v^{\mu}+F_3v'^{\mu})u_{\Lambda_b}
(v)\\
&&-\bar{u}_{\Lambda_c}(v')(G_1\gamma^{\mu}+G_2v^{\mu}+G_3v'^{\mu})\gamma^5
u_{\Lambda_b}(v)~,
\end{array}
\label{eq:form}
\end{equation}
where $v$ and $v'$ denote the four-velocities of $\Lambda_b$ and 
$\Lambda_c$ respectively.  
Within the framework of HQET, the classification of the form factors is
simplified very much. To the order of both $1/m_c$ and $1/m_b$, they
are expressed as
\begin{equation}
\begin{array}{lll}
F_1&=&\displaystyle C(\mu)\xi(y)+\big(\frac{\bar{\Lambda}}{2m_c}
+\frac{\bar{\Lambda}}{2m_b}\big)[2\chi(y)+\xi(y)]~,\\[3mm]
G_1&=&\displaystyle C(\mu)\xi(y)+\big(\frac{\bar{\Lambda}}{2m_c}
+\frac{\bar{\Lambda}}{2m_b}\big)[2\chi(y)+\frac{y-1}{y+1}\xi(y)]~,\\[3mm]
F_2&=&\displaystyle G_2=-\frac{\bar{\Lambda}}{m_c(y+1)}\xi(y)~,\\[3mm]
F_3&=&\displaystyle -G_3=-\frac{\bar{\Lambda}}{m_b(y+1)}\xi(y)~\\[3mm]
\end{array}
\end{equation} 
with the perturbative QCD coefficient in the leading logarithmic
approximation
\begin{equation}
C(\mu)=\left[\frac{\alpha_s(m_b)}{\alpha_s(m_c)}\right]^{\frac{6}{25}}
\left[\frac{\alpha_s(m_c)}{\alpha_s(\mu)}\right]^{a_L(y)}~,\\[3mm]
\end{equation}
where
\begin{equation}
a_L(y)=\frac{8}{27}[yr(y)-1]~, ~~~r(y)=\frac{1}{\sqrt{y^2-1}}
\ln(y+\sqrt{y^2-1}).
\end{equation}
$\xi$ and $\chi$ are the so-called leading and subleading Isgur-Wise function
respectively.  And the mass parameter
$\bar{\Lambda}$ is defined as follows
\begin{equation}
\bar{\Lambda}=m_{\Lambda_Q}-m_Q~.
\end{equation} 
\par
\vspace{1.0cm}
\indent
By QCD sum rules, $\xi$, $\chi$ and 
$\bar{\Lambda}$ have been obtained \cite{Dai}.
QCD sum rule is regarded as a nonperturbative method 
rooted in QCD itself \cite{Shifman}.
In a linear approximation, the leading Isgur-Wise function is fit as 
\begin{equation}
\xi(y)=1-\rho^2(y-1)~,~~~~~\rho^2=0.55\pm0.15~.
\label{eq:sumrule}
\end{equation}
On the other hand, the subleading Isgur-Wise function is 
negligibly small,
\begin{equation}
\chi(y)\simeq O(10^{-2})~.
\end{equation}
And the parameter $\bar{\Lambda}$ is determined to be
\begin{equation}
\bar{\Lambda}=0.79\pm 0.05 ~{\rm GeV}~.
\end{equation}
\par
\vspace{1.0cm}
   It is interesting to compare above QCD sum rule results with that of
large $N_c$.  Large $N_c$ limit is one of the most important and 
model-independent method of nonperturbative QCD in spite of the realistic
$N_c=3$ \cite{Hooft}.  
HQET in this limit, for the heavy baryon case, is 
often believed to be the heavy quark Skyrme model \cite{JenkinsL}.
The leading Isgur-Wise function is predicted as \cite{ManoharL} 
\begin{equation}
\xi (y)=0.99\exp [-1.3(y-1)]~. 
\label{eq:nc}
\end{equation}
The slope of this Isgur-Wise
function is steeper than that of the sum rule.  In the large $N_c$ limit,
the parameter $\bar{\Lambda}$ equals to the proton mass \cite{Chow}. 
In the following analysis, 
we take it as $\bar{\Lambda}\simeq 0.87$ GeV \cite{Chow}.  
This result is in agreement with that obtained by QCD sum rules.  
However, there is no Skyrme model calculation 
for the subleading Isgur-Wise function.\par  
We will assume that the subleading Isgur-Wise function is negligible
in the Skyrme model analysis.
In the following, we argue that this assumption is true 
in the large $N_c$ limit.
The subleading Isgur-Wise function $\chi(y)$ is defined by
\begin{eqnarray}
& &\langle\Lambda_c(v')|{\rm T}\bar{h}^{(c)}_{v'}\Gamma h^{(b)}_v
i\int d^4x\frac{1}{2m_Q}\bar{h}^{(Q)}_v(x)(iD)^2h^{(Q)}_v(x)
|\Lambda_b(v)\rangle\nonumber \\
&=& \frac{\bar{\Lambda}}{m_Q}\chi(y)
\bar{u}_{\Lambda_c}(v')\Gamma u_{\Lambda_b}(v)~,\\[3mm]
\end{eqnarray}
where $h^{(Q)}_v$ denotes the heavy quark field defined in the HQET with 
velocity $v$, and $\Gamma$ is some gamma matrix.  
$\chi(y)$ just measures 
the amplitude of the brown muck transfer through a strong
interaction, described by above matrix element, from the 
heavy quark which has a velocity change from $v$ to $v'$ due to the weak decay.
In the Hartree-Fock picture of large $N_c$ HQET, 
heavy baryon has $N_c-1$ light quarks.
Any $v\neq v^\prime ~~(y\neq 1)$ transition in fact is suppressed
when $N_c$ is large, because that involves changing the momenta of all
the light quarks inside the baryon.
In the limit $N_c\to\infty$, we expect $\chi(y\neq 1)=0$.
Furthermore, 
it is well-known that $\chi(1)=0$ due to Luke theorem \cite{LukeL}. 
Therefore, we get that $\chi(y)$ vanishes in the large $N_c$ limit.
Although the above argument makes our assumption for $\chi(y)$ reasonable,
it should be noted that there is still a subtle point which distinguishes
the Skyrme model from the large $N_c$ limit.
The point is that for heavy baryon weak decay form factors, 
the Skyrme model result is not exactly identical to that of large $N_c$.
Consider the leading Isgur-Wise function,
our large $N_c$ argument for $\chi(y)$ also applies to $\xi(y)$,
that is $\xi(y\neq 1)=0$.
Because $\xi(1)=1$, we expect that the leading Isgur-Wise function is 
$\delta$-function like, $\xi(y)\sim\delta(y-1)$ in the large $N_c$ limit.
However, this result in principle agrees with 
that of Skyrme model \cite{ManoharL} $\xi(y)\sim{\rm{exp}}[-N_c^{3/2}(y-1)]$, 
if $N_c$ is taken to be $\infty$.
The so called Skyrme model result Eq.(\ref{eq:nc}) can be obtained 
by taking $N_c=3$.
Therefore, $\chi(y)=0$ can be understood as the result of 
large $N_c$ limit for the Skyrme model.
%%%%%%%%%%%%%%%%%%%%%%%%%%%%%%%%%%%%%%%%%%%%%%%%%%%%%%%%%%%%%%%%%%%%%%%%%%%
\section{Decay Rates, Distributions and Asymmetry Parameters}
%%%%%%%%%%%%%%%%%%%%%%%%%%%%%%%%%%%%%%%%%%%%%%%%%%%%%%%%%%%%%%%%%%%%%%%%%%%
With the knowledge of the form factors from QCD sum rule and large $N_c$ limit,
we can calculate the rates, distributions and various asymmetry parameters 
for the $\Lambda_b\to\Lambda_c$ semileptonic decay.
The standard expressions for these ovservables are given in Ref.
\cite{Koerner} in terms of helicity amplitudes.
The process $\Lambda_b\to\Lambda_c l\bar{\nu}$ is considered
as a two-successive decay
$\Lambda_b \rightarrow \Lambda_c +W_{\rm {off-shell}}$, 
$W_{\rm{off-shell}} \rightarrow l+\nu$.
Let $\epsilon^\mu_{\lambda_W}$ be the polarization vector
of $W_{\rm{off-shell}}$, where $\lambda_W$ denotes the
helicity state.
Longitudinal state corresponds to $\lambda_W=0$, whereas
transverse state, $\lambda_W=\pm 1$. 
The helicity amplitudes are defined by
\begin{equation}
H^{V(A)}_{\pm \frac{1}{2} \lambda_W}
=\epsilon^\mu_{\lambda_W} 
 \langle\Lambda_c(v^\prime;\pm \frac{1}{2})|J^{V(A)}_\mu|
 \Lambda_b(v)\rangle~,
\end{equation}
where $J^{V(A)}$ stands for the vector(axial vector) current,
and $\pm\frac{1}{2}$ in the subscript is the helicity of
the daughter baryon $\Lambda_c$.
They can be expressed by the form factors,
\begin{eqnarray}
  \sqrt{q^2}H^{V,A}_{\frac{1}{2} 0}
  &=&\sqrt{2M_{\Lambda_b}M_{\Lambda_c}(y\mp 1)}
       \{(M_{\Lambda_b}\pm M_{\Lambda_c})(F_1,G_1)\nonumber\\
  & &\pm M_{\Lambda_c}(y\pm 1)(F_2,G_2)
        \pm M_{\Lambda_b}(y\pm 1)(F_3,G_3)\}~,\nonumber\\
  H^{V,A}_{\frac{1}{2} 1}
  &=& -2\sqrt{M_{\Lambda_b}M_{\Lambda_c}(y\mp 1)}
      (F_1,G_1)~, 
\label{eq:H}
\end{eqnarray} 
where the upper(lower) sign is for the vector(axial vector)
current.
With the notation of the total helicity amplitude
$H_{\pm \frac{1}{2} \lambda_W}
  = H^V_{\pm\frac{1}{2}\lambda_W}
     -H^A_{\pm\frac{1}{2}\lambda_W}$, 
and the parity relation
$H^{V(A)}_{\mp \frac{1}{2}-\lambda_W}
  =(-)H^{V(A)}_{\pm\frac{1}{2}\lambda_W}$, 
the differential decay rate can be expressed as 
\begin{eqnarray}
  \frac{d\Gamma}{dy~ d\cos\theta}
  &=&\frac{G_F^2}{(2\pi)^3}|V_{cb}|^2
     q^2\sqrt{y^2-1}\frac{M_{\Lambda_c}^2}{M_{12\Lambda_b}}\big [
     \frac{3}{8}(1+\cos\theta)^2|H_{\frac{1}{2}1}|^2\nonumber\\
    & &+\frac{3}{8}(1-\cos\theta)^2|H_{-\frac{1}{2}-1}|^2
     +\frac{3}{4}\sin^2\theta(|H_{\frac{1}{2}0}|^2
     +|H_{-\frac{1}{2}0}|^2)\big ], 
\label{eq:a}
\end{eqnarray}
where $\theta$ is the angle between $P_{\Lambda_c}$ and
$p_l$ measured in the $W_{\rm{off-shell}}$ rest frame.
The $y$ distribution of the decay rate is obtained by the integration
over $\cos\theta$,
\begin{eqnarray}
   \frac{d\Gamma}{dy}
      &=&\frac{G_F^2}{(2\pi)^3}|V_{cb}|^2q^2\sqrt{y^2-1}
         \frac{M_{\Lambda_c}^2}{12M_{\Lambda_b}}
         \Big[|H_{\frac{1}{2}1}|^2+|H_{-\frac{1}{2}-1}|^2
          +|H_{\frac{1}{2}0}|^2+|H_{-\frac{1}{2}0}|^2\Big]\nonumber\\
      &=&\frac{d\Gamma_{T+}}{dy}+\frac{d\Gamma_{T-}}{dy}
        +\frac{d\Gamma_{L+}}{dy}+\frac{d\Gamma_{L-}}{dy}~,
\label{eq:dgdy}
\end{eqnarray}
where $T_\pm$, $L_\pm$ are defined as the transerve and
longitudinal contribution to the decay rate with $\pm$ final
baryon helicity, respectively.
\\ \indent
The numerical results are obtained by inputting the form factors 
discussed in last section.
We have taken $m_c=1.44~{\rm GeV}$, $m_b=4.83~{\rm GeV}$, $\mu=0.47~{\rm GeV}$ 
and $|V_{cb}|=0.04$.
The partial decay distributions are
plotted as a function of $y$ in 
Fig.1 and Fig.2
for both QCD sum rule and large $N_c$ predictions.
It is easy to see
the dominance of $\frac{d\Gamma_{T-}}{dy}$ and $\frac{d\Gamma_{L-}}{dy}$
over other plus helicity components.
As discussed in Ref. \cite{Kroll}, this is due to the left-handed V-A
current.
From Fig.2, 
we can see that the discrepancy of different model gets larger 
as $y$ goes larger.
As a result of the fact that the slope of the Isgur-Wise function
of large $N_c$ is steeper than that of QCD sum rule,
the decay distributions on $y$ predicted by large $N_c$ are smaller than
that of QCD sum rule explicitly when $y\gtrsim 1.1$ .\par
It is experimentally useful to calculate
the lepton energy distribution. 
We obtain
\footnote{Our results for 
 $\frac{d\Gamma_{L_\pm}}{dE_l}$ 
are different from that given in Ref. \cite{IvanovL}.},
\begin{eqnarray}
  \frac{d\Gamma}{dE_l}&=&
    \frac{G_F^2}{(2\pi)^3}|V_{cb}|^2\frac{M^2_{\Lambda_c}}{8}
    \int_{y_{min}(E_l)}^{y_{max}}dy(y_{max}-y)\Big[
       (1+\cos\theta)^2|H_{\frac{1}{2}1}|^2            \nonumber\\
  & &
      +(1-\cos\theta)^2|H_{-\frac{1}{2}-1}|^2     
      +2\sin^2\theta\big(|H_{\frac{1}{2}0}|^2
                       +|H_{-\frac{1}{2}0}|^2
                        \big)\Big]                      \nonumber\\
  &\equiv&\frac{d\Gamma_{T_+}}{dE_l}
         +\frac{d\Gamma_{T_-}}{dE_l}
         +\frac{d\Gamma_{L_+}}{dE_l}
         +\frac{d\Gamma_{L_-}}{dE_l}
\label{eq:b}
\end{eqnarray}
where
\begin{eqnarray}
   \cos\theta &=&
       \frac{E^{max}_l-2E_l+M_{\Lambda_c}(y_{max}-y)}
            {M_{\Lambda_c}\sqrt{y^2-1}}              \nonumber\\
   E^{max}_l &=& 
     \frac{M^2_{\Lambda_b}-M^2_{\Lambda_c}}{2M_{\Lambda_b}}\nonumber\\
   y_{max} &=& 
     \frac{M^2_{\Lambda_b}+M^2_{\Lambda_c}}{2M_{\Lambda_b}M_{\Lambda_c}}
          \nonumber\\
   y_{min}(E_l) &=& y_{max}
     -2\frac{E_l}{M_{\Lambda_c}}\frac{E^{max}_l-E_l}
           {M_{\Lambda_b}-2E_l}.
\end{eqnarray}
The lepton energy spectrums of the decay rates are given 
in Fig.3 and Fig.4
for the QCD sum rule and large $N_c$ Isgur-Wise function.
As in the case of $y$ distribution, the helicity minus components
dominate the plus ones.
And the decay distributions on $E_l$ by large $N_c$ are always
smaller than that by QCD sum rule.
The decay rates are obtained from Eq.(\ref{eq:dgdy}) by integrating over $y$,
or from Eq.(\ref{eq:b}) by integrating over $E_l$.
The numerical results for the partial decay rates into given helicity
states are listed in Table 1, 
where the quark model results \cite{Kroll} are also listed for comparison.
The total decay rate is obtained by summing them up,
\begin{equation}
  \Gamma=6.17\times10^{-14}~\rm{GeV}~,~
  {\rm Br.}
    {\it(\Lambda_b\to\Lambda_c l \bar{\nu})}=11.5\%
    \times
    \Big(\frac{\tau({\it\Lambda_b})}{1.23\times10^{-12}~{\rm sec}}\Big)
\end{equation}
for QCD sum rule linear fitting, and
\begin{equation}
\Gamma=4.51\times10^{-14}~\rm{GeV}~,~
  {\rm Br.}
   {\it(\Lambda_b\to\Lambda_c l \bar{\nu})}=8.43\%
   \times
   \Big(\frac{\tau({\it\Lambda_b})}{1.23\times10^{-12}~{\rm sec}}\Big)
\end{equation}
for large $N_c$ approximation.
The quark model results given in Ref. \cite{Kroll} are
$\Gamma=4.28\times10^{-14}~{\rm GeV}$ and ${\rm Br.}=7.99\%$
for
$\tau(\Lambda_b)=1.23\times 10^{-12}~{\rm sec}$.
The QCD sum rule predicts larger decay branching ratio than large $N_c$ 
model, as we have expected.
We also see that both QCD sum rule and large $N_c$ predicts larger 
results for the decay than quark model of Ref. \cite{Kroll}.
Up to the leading order, we have
$\Gamma=5.52\times 10^{-14}~{\rm GeV}$
for QCD sum rule and
$\Gamma=4.00\times 10^{-14}~{\rm GeV}$
for large $N_c$ limit.
It means that $1/m_c$ and $1/m_b$ corrections yield about $11\%$
enhancement for the total decay rate.
\\[10mm]
\indent
Now let's turn to the various asymmetry paremeters.
The polarization effects in the process 
$\Lambda_b\to\Lambda_c$ are revealed
in various angular distributions.
First, from Eq.(\ref{eq:a}), the polar angle distribution is
\begin{equation}
  \frac{d\Gamma}{dy~ d\cos\theta}\propto
    1+2\alpha^\prime\cos\theta+\alpha^{\prime\prime}cos^2\theta,
\end{equation}
where $\alpha^\prime$ and $\alpha^{\prime\prime}$ are asymmetry parameters
which can be expressed as
\begin{eqnarray}
  \alpha^\prime&=&\frac{|H_{1/2~1}|^2-|H_{-1/2-1}|^2}
      {|H_{1/2~1}|^2+|H_{-1/2-1}|^2+2(|H_{1/2~0}|^2+|H_{-1/2~0}|^2)}
  \label{eq:alpha1} \\
  \alpha^{\prime\prime}&=& \frac
     {|H_{1/2~1}|^2+|H_{-1/2-1}|^2-2(|H_{1/2~0}|^2+|H_{-1/2~0}|^2)}
     {|H_{1/2~1}|^2+|H_{-1/2-1}|^2+2(|H_{1/2~0}|^2+|H_{-1/2~0}|^2)}~.
  \label{eq:alpha2}
\end{eqnarray}
There are other asymmetry parameters if 
the successive hadronic cascade decay 
$\Lambda_c\rightarrow a+b$ where $a$ and $b$ are some hadrons, are considered.
Two more angles are involved, $\Theta_\Lambda$ and $\chi$.
$\Theta_\Lambda$ is the angle beween $\Lambda_c$'s momentum 
in $\Lambda_b$ rest frame and
$a$'s momentum in $\Lambda_c$ rest frame assuming $J_a=\frac{1}{2}$ 
and $J_b=0$.
$\chi$ is the relative azimuthal angle between the decay planes 
defined by $l,~\nu$ and $a,~b$.
$\Theta_\Lambda$ and $\chi$ distributions of the decay are 
 \cite{Koerner}
\begin{equation}
   \frac{d\Gamma}{dy~d\cos\Theta_\Lambda}\propto
   1+\alpha\alpha_\Lambda\cos\Theta_\Lambda~~{\rm and}~~
   \frac{d\Gamma}{dy~d\chi}\propto
   1-\frac{3\pi^2}{32\sqrt{2}}\gamma\alpha_\Lambda\cos\chi,
\end{equation}
where $\alpha_\Lambda$ is the asymmetry parameter in the 
$\Lambda_c$ hadronic decay. 
In this case, the related asymmetry parametes, 
$\alpha$ and $\gamma$ are given by
\begin{equation}
  \alpha=\frac
   {|H_{1/2~1}|^2-|H_{-1/2-1}|^2+|H_{1/2~0}|^2-|H_{-1/2~0}|^2}
   {|H_{1/2~1}|^2+|H_{-1/2-1}|^2+|H_{1/2~0}|^2+|H_{-1/2~0}|^2},
\end{equation}
\begin{equation}
   \gamma=\frac
    {2\rm{Re}(H_{-1/2~0}H^*_{1/2~1}+H_{1/2~0}H^*_{-1/2-1})}
    {|H_{1/2~1}|^2+|H_{-1/2-1}|^2+|H_{1/2~0}|^2+|H_{-1/2~0}|^2}.
\end{equation}
When the $\Lambda_b$ polarization is further considered,
additional asymmetry parameters can be introduced.
The new decay angles related are $\Theta_P$ and $\chi_P$.
$\Theta_P$ is the angle between $\Lambda_b$ polarization and $\Lambda_c$ 
momentum, and $\chi_P$ is the azimuthal angle between the plane of
$\Lambda_b$ polarization, $\Lambda_c$ momentum and that of $a$, $b$'s momenta.
The decay distributions are \cite{Koerner}
\begin{equation}
   \frac{d\Gamma_{pol}}{dy~d\cos\Theta_P}\propto
   1-\alpha_PP\cos\Theta_P ,~~{\rm and}~~
   \frac{d\Gamma_{pol}}{dy~d\chi_P}\propto
   1-\frac{\pi^2}{16}P\alpha_\Lambda\gamma_P\cos\chi_P,
\end{equation}
where $P$ is the degree of polarization of $\Lambda_b$. 
The asymmetry parameters $\alpha_P$ and $\gamma_P$ are
\begin{equation}
   \alpha_P=\frac
     {|H_{1/2~1}|^2-|H_{-1/2-1}|^2-|H_{1/2~0}|^2+|H_{-1/2~0}|^2} 
     {|H_{1/2~1}|^2+|H_{-1/2-1}|^2+|H_{1/2~0}|^2+|H_{-1/2~0}|^2},
\end{equation}
\begin{equation}
   \gamma_P=\frac{2\rm{Re}(H_{1/2~0}H^*_{-1/2~0})}
     {|H_{1/2~1}|^2+|H_{-1/2-1}|^2+|H_{1/2~0}|^2+|H_{-1/2~0}|^2}~.
\end{equation}
These asymmetry parameters are functions of $y$.
On averaging over $y$, the numerators and denominators are
integrated separately with proper weight, 
$(y_{max}-y)\sqrt{y^2-1}$.\\ \indent
Our numerical results on the mean values of the asymmetry parameters
are listed in Table 2 where both QCD sum rule and 
large $N_c$ results are given.
Note that these results include $1/m_c$ and $1/m_b$ corrections.
The quark model results \cite{Kroll} are also listed for comparison.
Note that all the results include $1/m_c$ and $1/m_b$ corrections.

%%%%%%%%%%%%%%%%%%%%%%%%%%%%%%%%%%%%%%%%%%%%%%%%%%%%%%%%%%%%%%%%%%%
\section{Nonleptonic decays}
%%%%%%%%%%%%%%%%%%%%%%%%%%%%%%%%%%%%%%%%%%%%%%%%%%%%%%%%%%%%%%%%%%%
In this section, we will consider the two-body nonleptonic decay modes
$\Lambda_b\rightarrow\Lambda_c\pi(\rho)$ and 
$\Lambda_b\rightarrow\Lambda_c K^{(*)}$.  
All these decays involve
external $W$-emission diagrams which can be analyzed by using the factorization
approximation \cite{Bauer,Neubert2}.  
$\Lambda_b\rightarrow\Lambda_c\pi(\rho)$ also get
contributions from internal $W$-emission which, however, is non-factorizable and
is difficult to calculate reliably.  
We will simply neglect this contribution in the analysis.  
Penguin diagrams do not contribute to these decays.
There are contributions from 
W-exchange diagrams in both $\Lambda_b\rightarrow\Lambda_c\pi(\rho)$ and 
$\Lambda_b\rightarrow\Lambda_c K^{(*)}$ channels.  Because such diagrams are 
suppressed by the possibility of the two valence quark lines meeting in the 
region of $1/M_W$, they are neglected.  This is also justified by a detailed 
quark model analysis for b-baryon nonleptonic decays [20].  In short, the 
decays will be analyzed by using factorization assumption which is expected to
be good except for the $\Lambda_c\pi(\rho)$ channels. 
In a recent study \cite{Rusetsky}, the non-factorizable effect in decay 
$\Lambda_b\rightarrow\Lambda_c\pi$ has been estimated.  The total 
non-factorizable contribution is about 30\% of the factorizable one.  Although
it is indeed sizable, the factorizable effect is still dominant. 
\par
After factorization, the amplitude of the process can be 
expressed by the product of two matrix elements to which the form factors
given in Sec.II can be applied just as in the case of the semileptonic decay.
With the definition of the $\Lambda_b\to\Lambda_c$ matrix element
Eq.(\ref{eq:form}), 
the widths of the decays into pseudoscalar meson(P)
and vector meson(V)
can be easily calculated:
\begin{equation}
\Gamma(P)=\frac{G_F^2}{2\pi}|V_{cb}V_{ij}|^2f_P^2a_1^2M_{\Lambda_b}^3
%  \sqrt{[(1+r)^2-t_P^2][(1-r)^2-t_P^2]}\nonumber\\
% & & \big\{[(1+r)^2-t_P^2]|A|^2+[(1-r)^2-t_P^2]|B|^2\big\}
  \frac{r^2\kappa_P}{(1-\kappa_P^2)^2}\big[|A|^2+\kappa_P^2|B|^2\big]~,
\end{equation}
where
\begin{eqnarray}
A&=&(1-r)\Big[F_1+\frac{1+r}{2}\big(F_2+\frac{F_3}{r}\big)\Big]~,\nonumber\\
B&=&(1+r)\Big[G_1-\frac{1-r}{2}\big(G_2+\frac{G_3}{r}\big)\Big]~,\nonumber\\
\kappa_{P(V)}&=&\sqrt{\frac{(1-r)^2-t_{P(V)}^2}{(1+r)^2-t_{P(V)}^2}}~,
\end{eqnarray}
and

\begin{equation}
 \Gamma(V)=\frac{G_F^2}{4\pi}|V_{cb}V_{ij}|^2a_1^2f_V^2 M_{\Lambda_b}^3
  \frac{r^2\kappa_V}{(1-\kappa_V^2)^2}\big[
  2t_V^2\big(\kappa_V^2|F_1|^2+|G_1|^2\big)+\kappa_V^2|C|^2+|D|^2\big]~,
\end{equation}
where
\begin{eqnarray}
  C&=&(1+r)F_1+\frac{2r}{1-\kappa_V^2}\big(F_2+\frac{F_3}{r}\big)~,\nonumber\\
  D&=&(1-r)G_1-\frac{2r\kappa_V^2}{1-\kappa_V^2}\big(
     G_2+\frac{G_3}{r}\big)~.
\end{eqnarray}
Here $V_{ij}$ denotes $V_{ud}$ for $\pi(\rho)$ and $V_{us}$ for $K^{(*)}$, 
and $r\equiv\frac{M_{\Lambda_c}}{M_{\Lambda_b}}$, 
$t_{P(V)}\equiv\frac{m_{P(V)}}{M_{\Lambda_b}}$
where $m_{P(V)}$ is the mass of the pseudoscalar (vector) meson.
And $y=\frac{1+r^2-t_{P(V)}^2}{2r}$.
$a_1$ is the QCD coefficient which is taken as a free parameter in the 
discussion of nonleptonic decays
\cite{Bauer,Neubert2}.

The above expressions are spin averaged results.
If we take into account the spin effects, 
the spin up-down asymmetry of $\Lambda_b$ is a good parameter to analyze
$\Lambda_b\to\Lambda_c$ nonleptonic decay.
In this case, the decay rates are
\cite{WiseL,Cheng}
\begin{equation}
 \Gamma(P)\propto 1+\alpha(P)({\bf S}_{\Lambda_c}+
  {\bf S}_{\Lambda_b})\cdot\hat{\bf p}_{\Lambda_c}~,
\end{equation}
where
\begin{equation}
 \alpha(P)=-\frac{2\kappa_P{\rm Re}(A^*B)}{|A|^2+\kappa_P^2|B|^2}~,
\end{equation}
and
\begin{equation}
 \Gamma(V)\propto 1+\alpha(V){\bf S}_{\Lambda_b}\cdot 
  \hat{\bf p}_{\Lambda_c}~,
\end{equation}
where
\begin{equation}
 \alpha(V)=\frac{\kappa_V{\rm Re}(8t_V^2F_1G_1^*-CD^*)}
   {2\big[2t_V^2(\kappa_V^2|F_1|^2+|G_1|^2)+\kappa_V^2|C|^2+|D|^2\big]}~.
\end{equation}
Here ${\bf S}_{\Lambda_Q}$ is the spin vector of $\Lambda_Q$ and
$\hat{{\bf p}}_{\Lambda_c}$ is the unit vector of momentum of
$\Lambda_c$.
$\alpha_{P(V)}$ is the spin up-down asymmetry parameter of $\Lambda_b$.
\par
The numerical results from QCD sum rule and large $N_c$ limit are given
in Table 3.
Because $y$ is near to 1.45, in the QCD sum rule case, we used the
nonlinear fitting of the Isgur-Wise function 
$\xi(y)=\big(\frac{2}{y+1}\big)^{0.5}
        \exp{\big[-0.8\frac{y-1}{y+1}\big]}$  \cite{Dai}.
The value of $a_1$ is taken to be 0.98 which is obtained from the decay
$B\to D\pi$ \cite{Neubert2}.
Various decay constants are taken as follows; 
$f_\rho=210~{\rm MeV}$, $f_K=158~{\rm MeV}$ and $f_{K^*}=214~{\rm MeV}$.
The quark model results of Ref. \cite{Cheng} are also listed for comparison.
%%%%%%%%%%%%%%%%%%%%%%%%%%%%%%%%%%%%%%%%%%%%%%%%%%%%%%%%%%%%%%%%%%%
\section{Summary}
%%%%%%%%%%%%%%%%%%%%%%%%%%%%%%%%%%%%%%%%%%%%%%%%%%%%%%%%%%%%%%%%%%%
We have analyzed the $\Lambda_b\to\Lambda_c$ semileptonic decay
in the framework of HQET to the order of $1/m_c$ and $1/m_b$.
The predictions for the Isgur-Wise functions and mass parameters 
from QCD sum rule and large $N_c$ method are used.
In the large $N_c$ limit, we argued that the subleading
Isgur-Wise function vanishes.
The decay rates, distributions and varuous asymmetry parameters
are calculated numerically.
Some of the $\Lambda_b$ nonleptonic decays are also calculated.
The numerical results can be checked by the experiments 
in the near future.
\\[10mm]
\begin{center}
{\large\bf Acknowledgments}\\[10mm]
\end{center}\par
We would like to thank Jungil Lee for helpful discussion.
This work was supported in part by the KOSEF through SRC program, 
and in part by the Korean Ministry of Education, 
Project No. BSRI-97-2418.
After finishing this work, we noticed a paper \cite{Giri} which calculated
the nonleptonic $\Lambda_b$ decays in the large $N_c$ and heavy quark limits.
\newpage
%%%%%%%%%%%%%%%%%%%%%%%%%%%%%%%%%%%%%%%%%%%%%%%%%%%%%%%%%%%%%%%%%%%%%%%%%%%%%%
%%%%%%%%%%%%%%%%%%%%%%%%%%%%%%%%%%%%%%%%%%%%%%%%%%%%%%%%%%%%%%%%%%%%%%%%%%%%%%
%%%%%%%%%%       Liu's file                
%%%%%%%%%%%%%%%%%%%%%%%%%%%%%%%%%%%%%%%%%%%%%%%%%%%%%%%%%%%%%%%%%%%%%%%%%%%%%%
%%%%%%%%%%%%%%%%%%%%%%%%%%%%%%%%%%%%%%%%%%%%%%%%%%%%%%%%%%%%%%%%%%%%%%%%%%%%%%

%%%%%%%%%%%%%%%%%%%%%%%%%%%%%%%%%%%%%%%%%%%%%%%%%%%%%%%%%%%%%%%%%%%%%%%%%%%%%%%%

\newpage
%%%%%%%%%%%%%%%%%%%%%%%%%%%%%%%%%%%%%%%%%%%%%%%%%%%%%
\begin{center} {\large\bf FIGURE CAPTIONS} \end{center}
%%%%%%%%%%%%%%%%%%%%%%%%%%%%%%%%%%%%%%%%%%%%%%%%%%%%%
\noindent
%=======
Fig~1.
%=======
$y$ distribution of the decay rates for (a) QCD sum rule and (b) large $N_c$ limit.
$\frac{d\Gamma_{T_\pm}}{dy}$ and $\frac{d\Gamma_{L_\pm}}{dy}$
are abbreviated as $T_\pm$ and $L_\pm$, respectively.\\
\vskip .3cm
\noindent
%=======
Fig~2.
%=======
Comparison of the two models in the helicity componential
$y$ distribution of the decay rates:
(a) $\frac{d\Gamma_{T_+}}{dy}$ (b) $\frac{d\Gamma_{T_-}}{dy}$ 
(c) $\frac{d\Gamma_{L_+}}{dy}$ (d) $\frac{d\Gamma_{L_-}}{dy}$ 
(e) $\frac{d\Gamma}{dy}=
\frac{d\Gamma_{T_+}}{dy}+\frac{d\Gamma_{T_-}}{dy}
+\frac{d\Gamma_{L_+}}{dy}+\frac{d\Gamma_{L_-}}{dy}$.\\
\vskip .3cm
\noindent
%=======
Fig~3.
%=======
Lepton energy distribution of the decay rates 
for (a) QCD sum rule and (b) large $N_c$ limit.\\
\vskip .3cm
\noindent
%=======
Fig~4.
%=======
Comparison of the two models in the helicity componential 
lepton energy distribution of the decay rates:
(a) $\frac{d\Gamma_{T_+}}{dE}$ (b) $\frac{d\Gamma_{T_-}}{dE}$
(c) $\frac{d\Gamma_{L_+}}{dE}$ (d) $\frac{d\Gamma_{L_-}}{dE}$
(e) $\frac{d\Gamma}{dE}=
\frac{d\Gamma_{T_+}}{dE}+\frac{d\Gamma_{T_-}}{dE}
+\frac{d\Gamma_{L_+}}{dE}+\frac{d\Gamma_{L_-}}{dE}$.\\

%%%%%%%%%%%%%%%%%%%%%%%%%%%%%%%%%%%%%%%%%%%%%%%%%%%%%
\begin{center}
{\large\bf TABLE CAPTIONS}
\end{center}
%%%%%%%%%%%%%%%%%%%%%%%%%%%%%%%%%%%%%%%%%%%%%%%%%%%%%
\noindent
%=======
Table 1.
%=======
The partial decay rates(in $10^{-14}~{\rm GeV}$).\\
\vskip .3cm
\noindent
%=======
Table 2.
%=======
Asymmetry parameters.\\
\vskip .3cm
\noindent
%=======
Table 3.
%=======
Numerical results for $\Lambda_b$ two-body nonleptonic decays. 
  The QCD coefficient $a_1$ is taken to be 0.98\cite{Neubert2}.\\ 
\newpage
%====================   FIGURE (y distribution) =============
%                    many figures enter!: y distribution
\begin{figure}
\vskip 2cm
\begin{center}
\epsfig{file=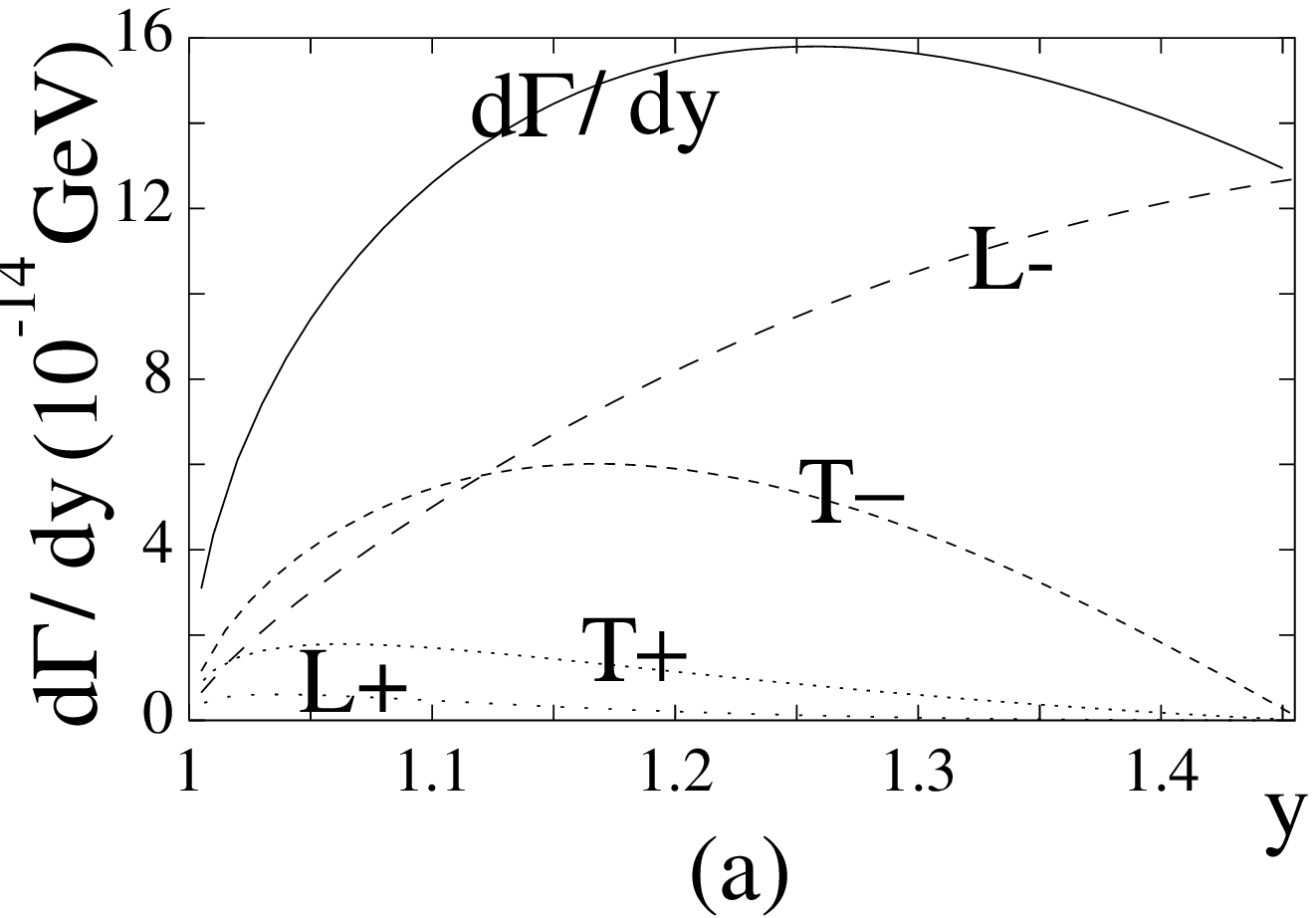, height=5cm}\\[5mm]
\epsfig{file=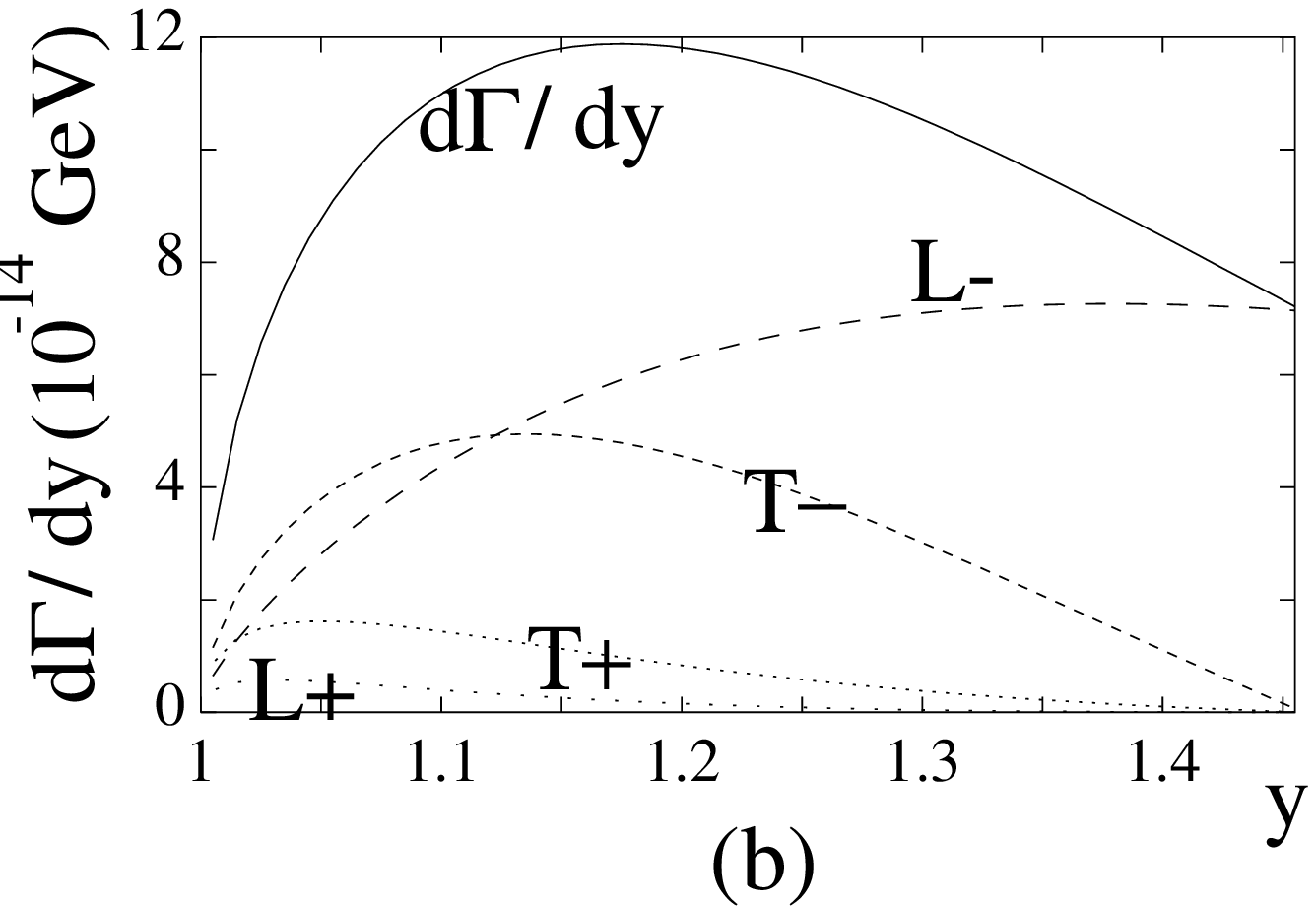, height=5cm}
\end{center}
\label{fig:yi}
\caption{}
\end{figure}

\begin{figure}
\begin{tabular}{cc}
\epsfig{file=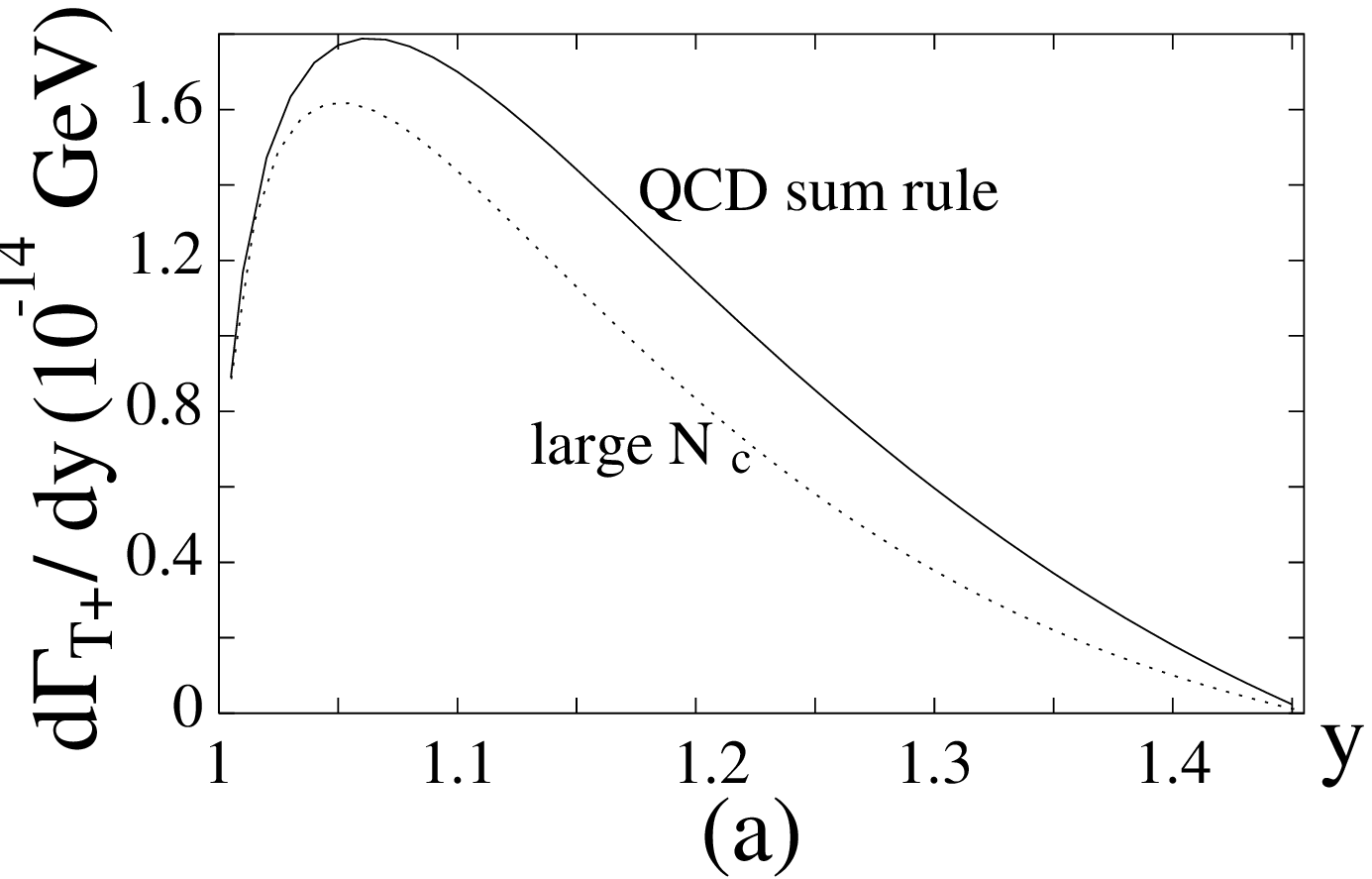, height=5cm}&
~~~\epsfig{file=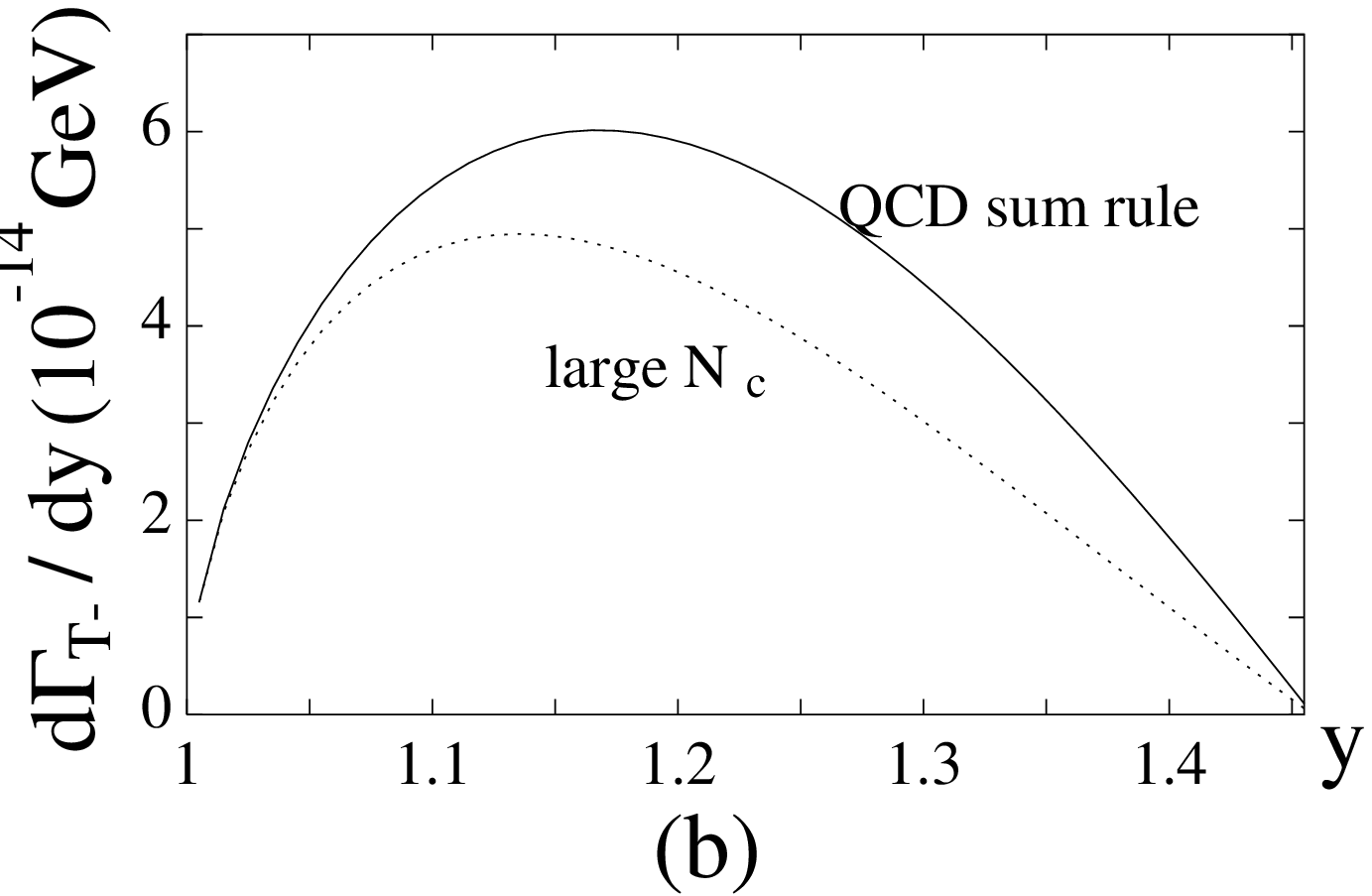, height=5cm}\\[5mm]
\epsfig{file=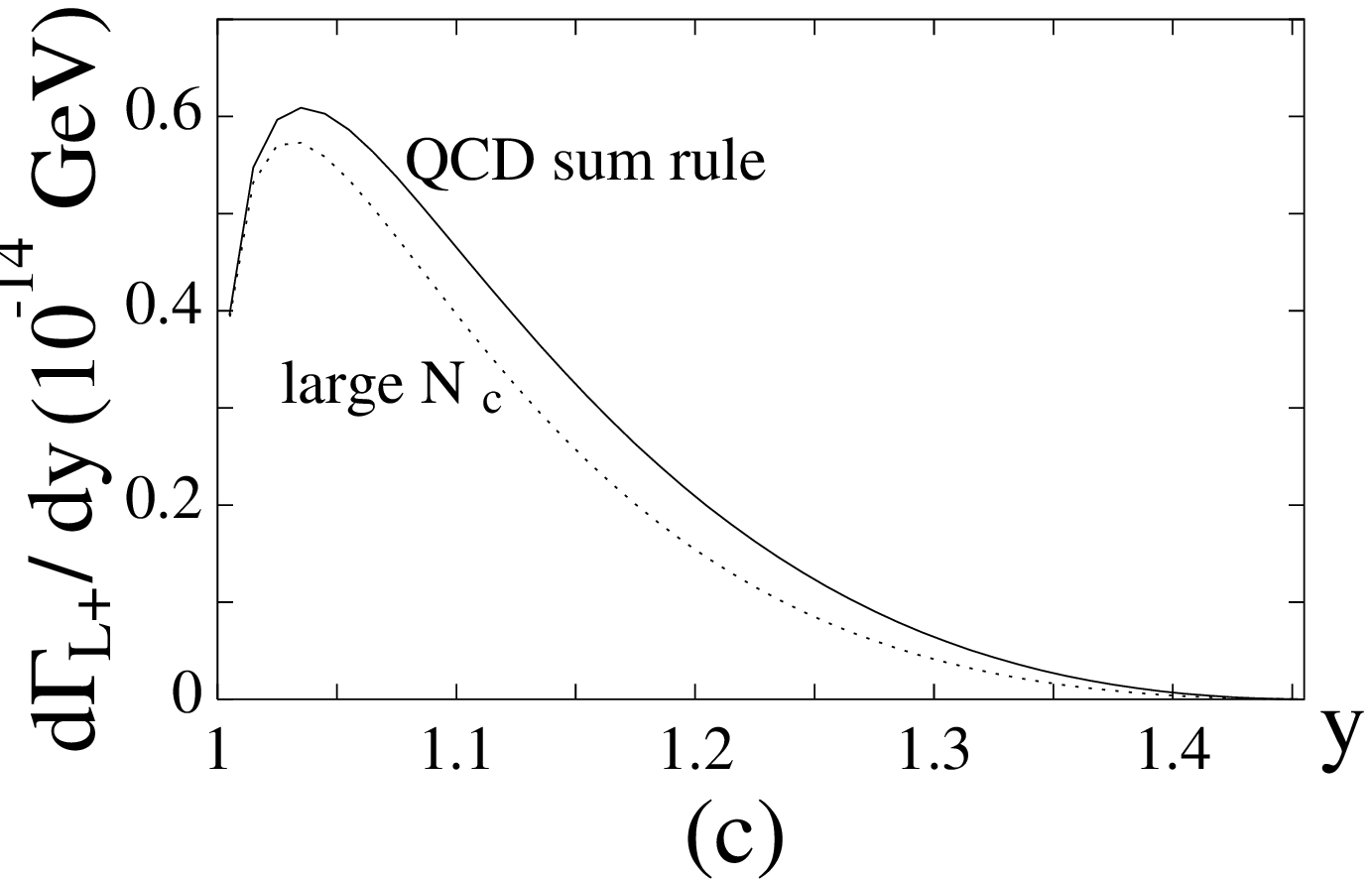, height=5cm}&
~~~\epsfig{file=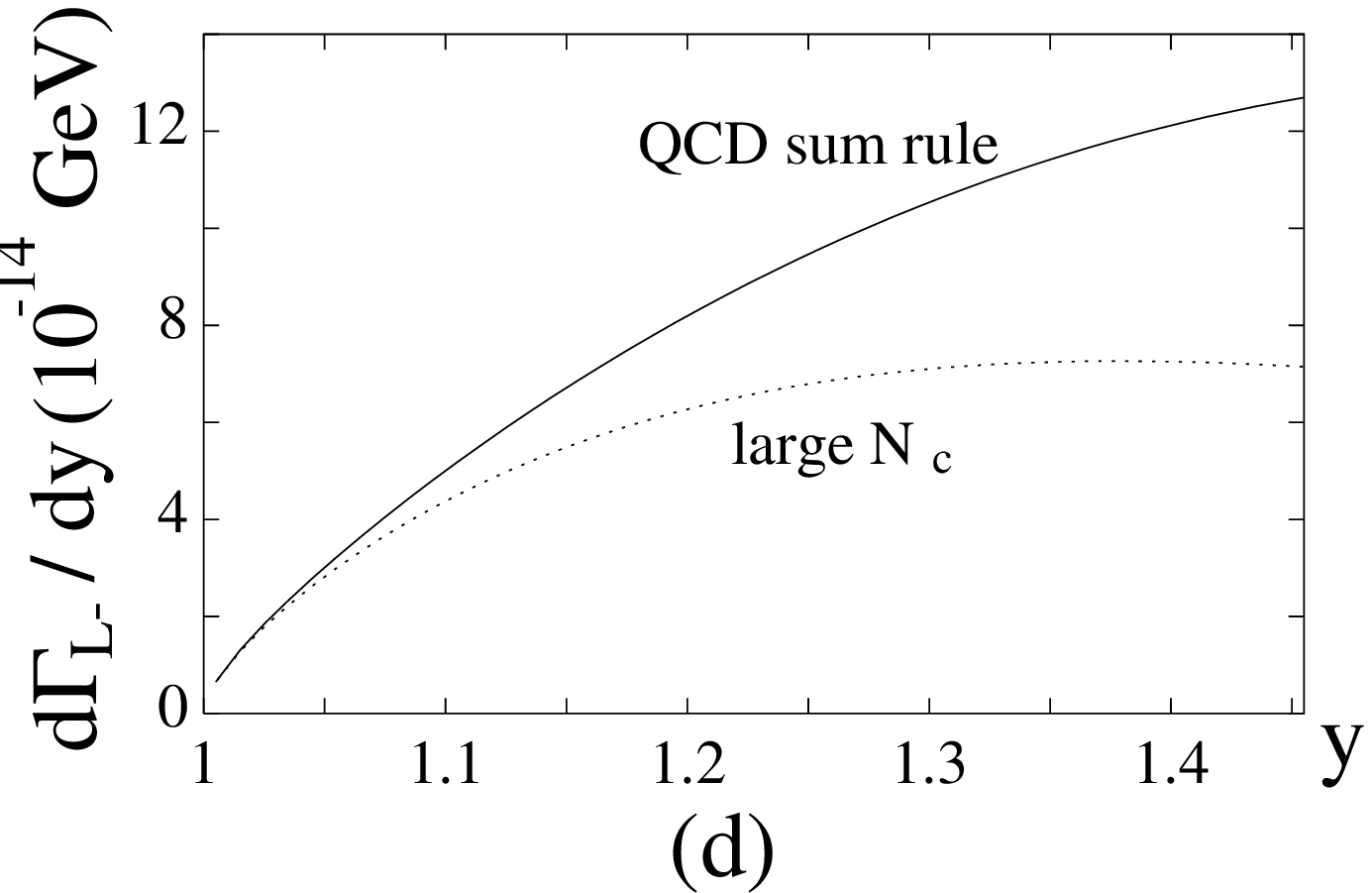, height=5cm}\\[5mm]
\epsfig{file=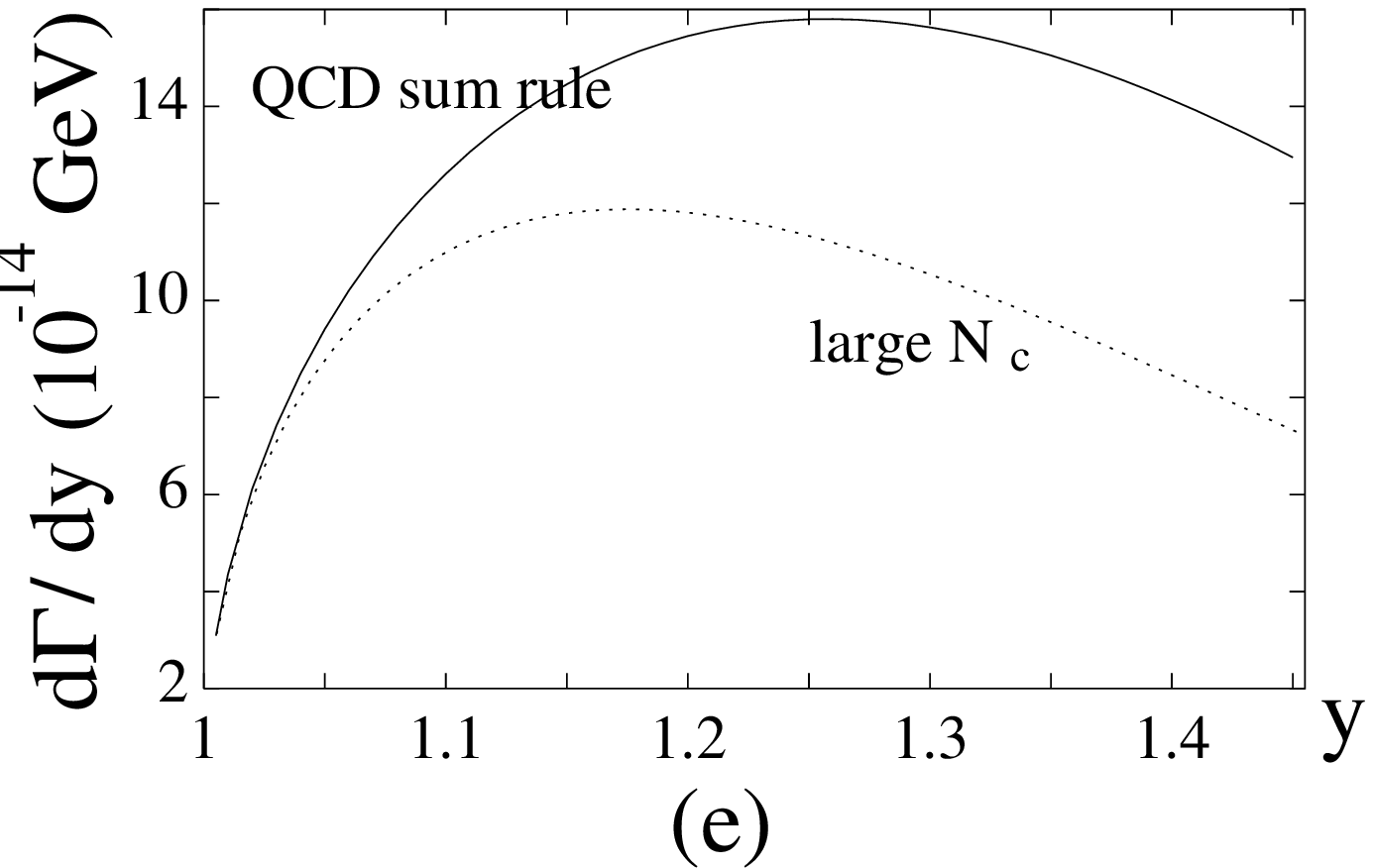, height=5cm}
\end{tabular}
\label{fig:yf}
\caption{}
\end{figure}
%==================   FIGURE(energy distribution)   ================

\begin{figure}
\vskip 2cm
\begin{center}
\epsfig{file=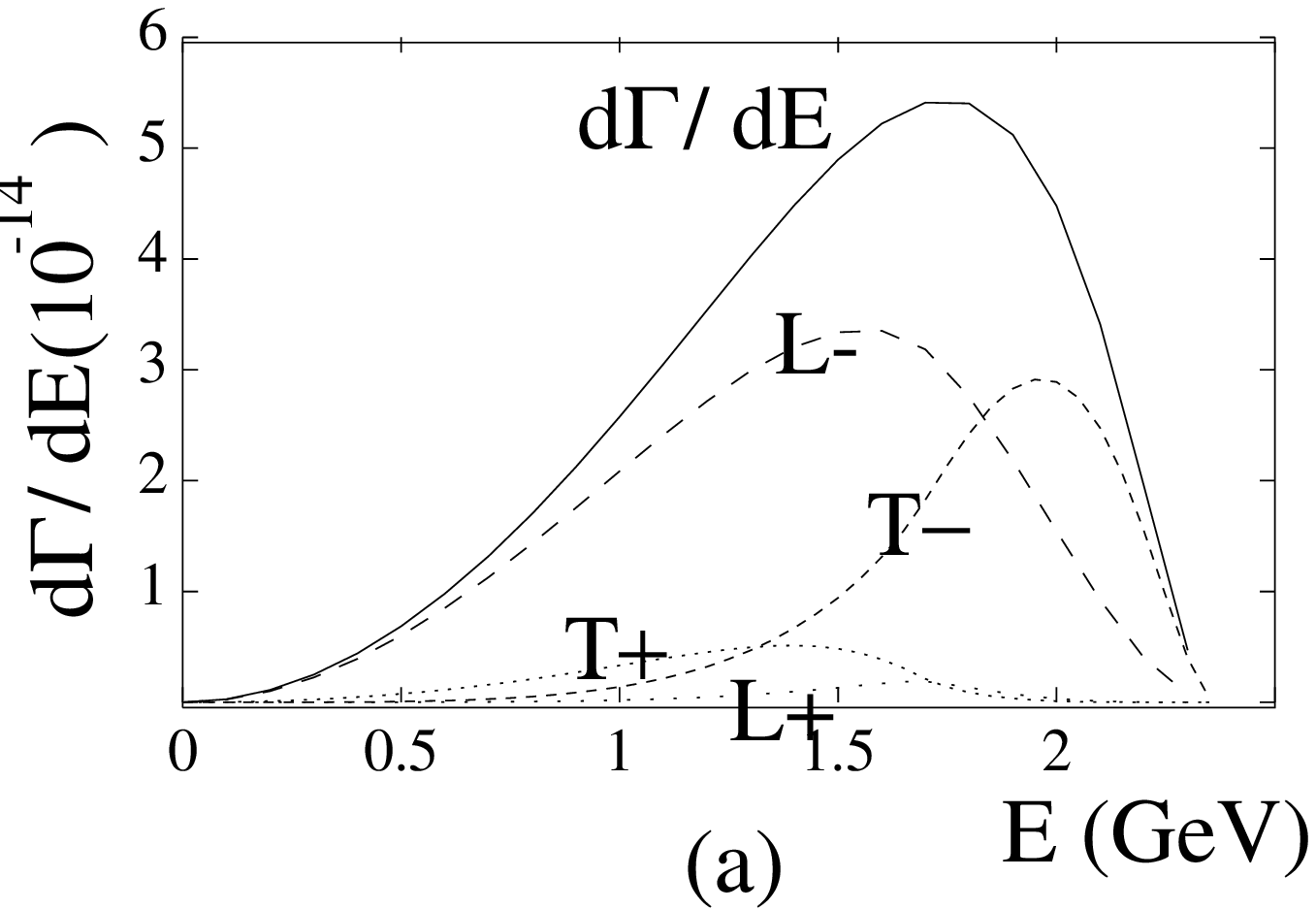, height=5cm}\\[5mm]
\epsfig{file=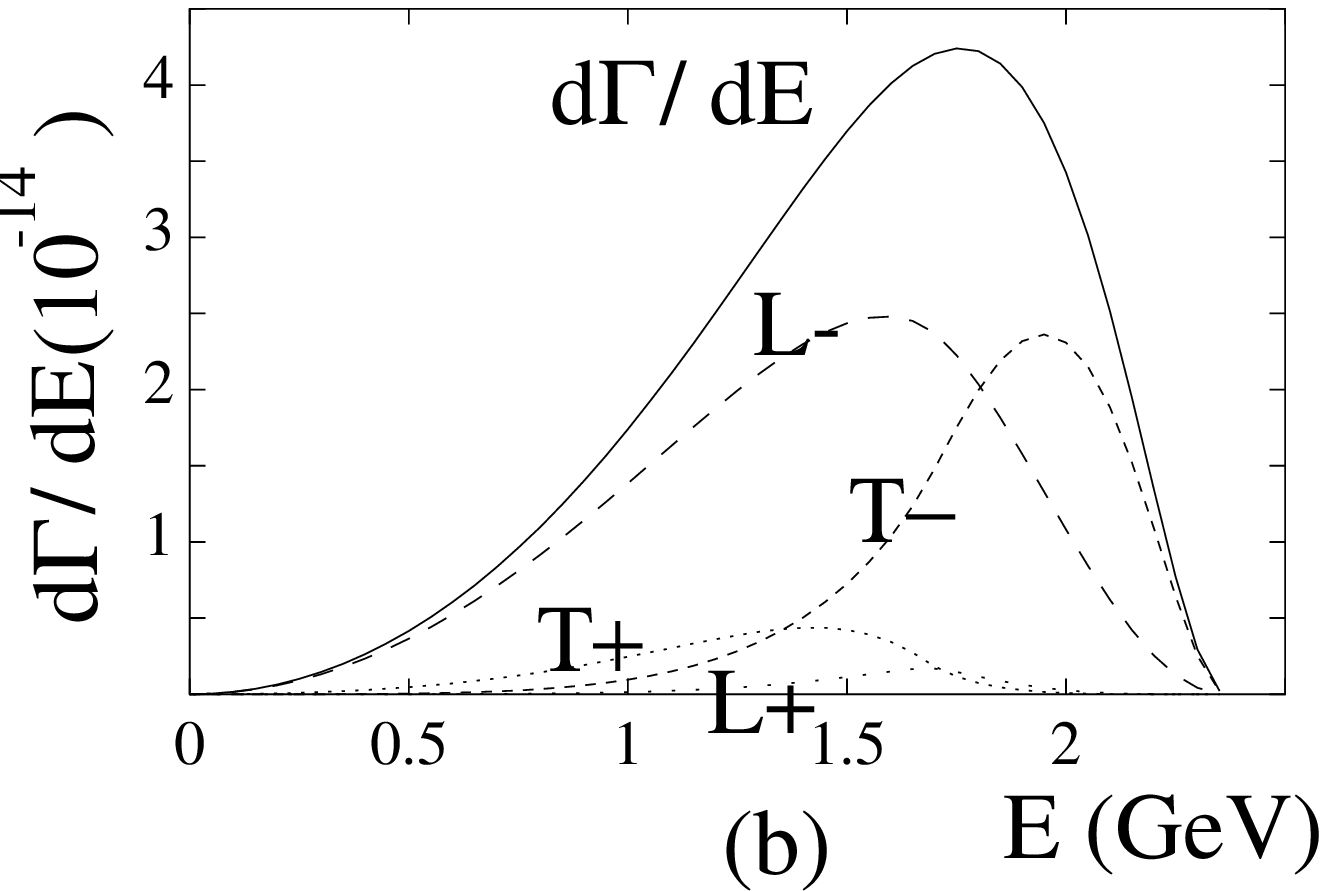, height=5cm}
\end{center}
\label{fig:ei}
\caption{}
\end{figure}

\begin{figure}
\begin{tabular}{cc}
\epsfig{file=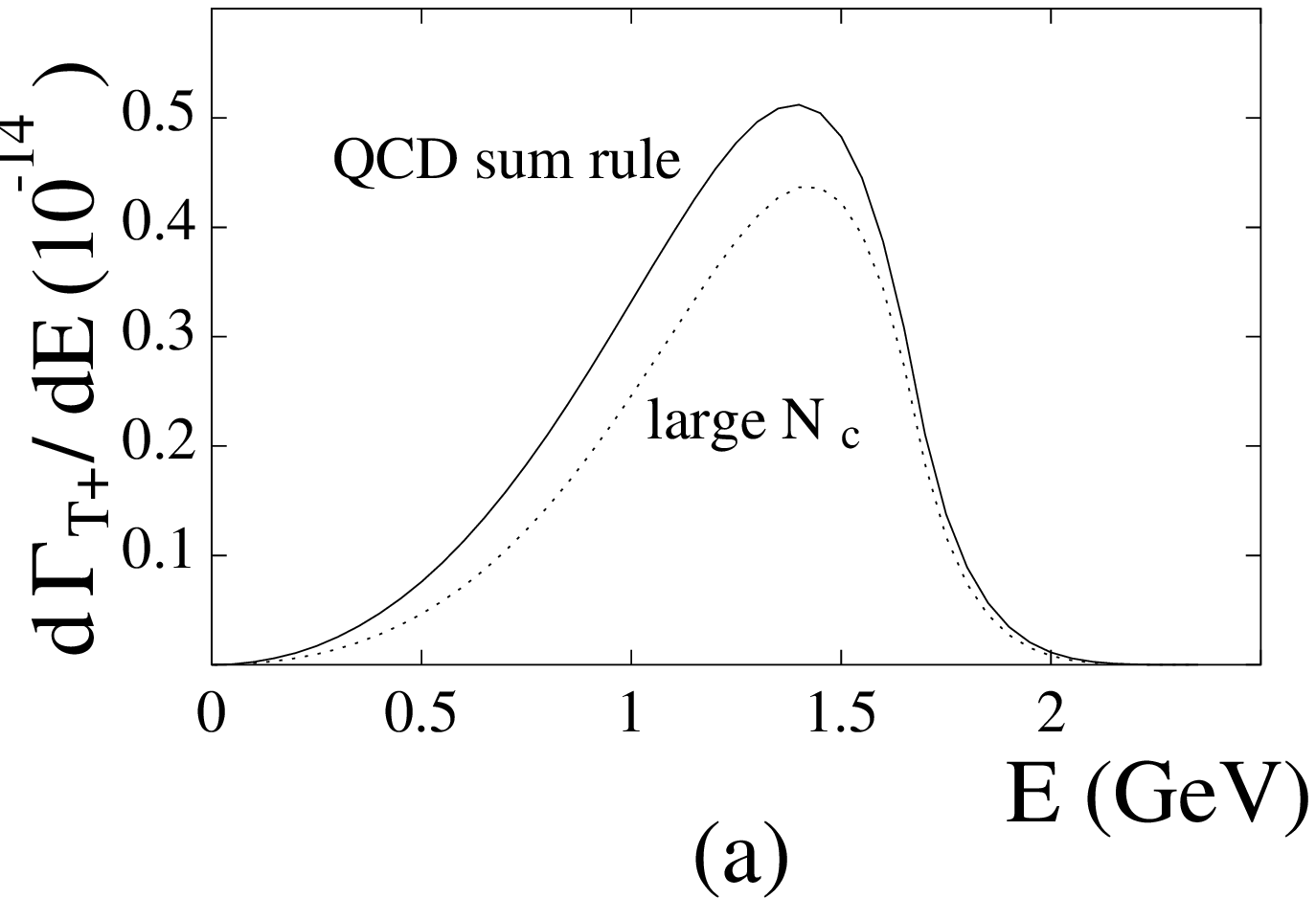, height=5cm}&
~~~\epsfig{file=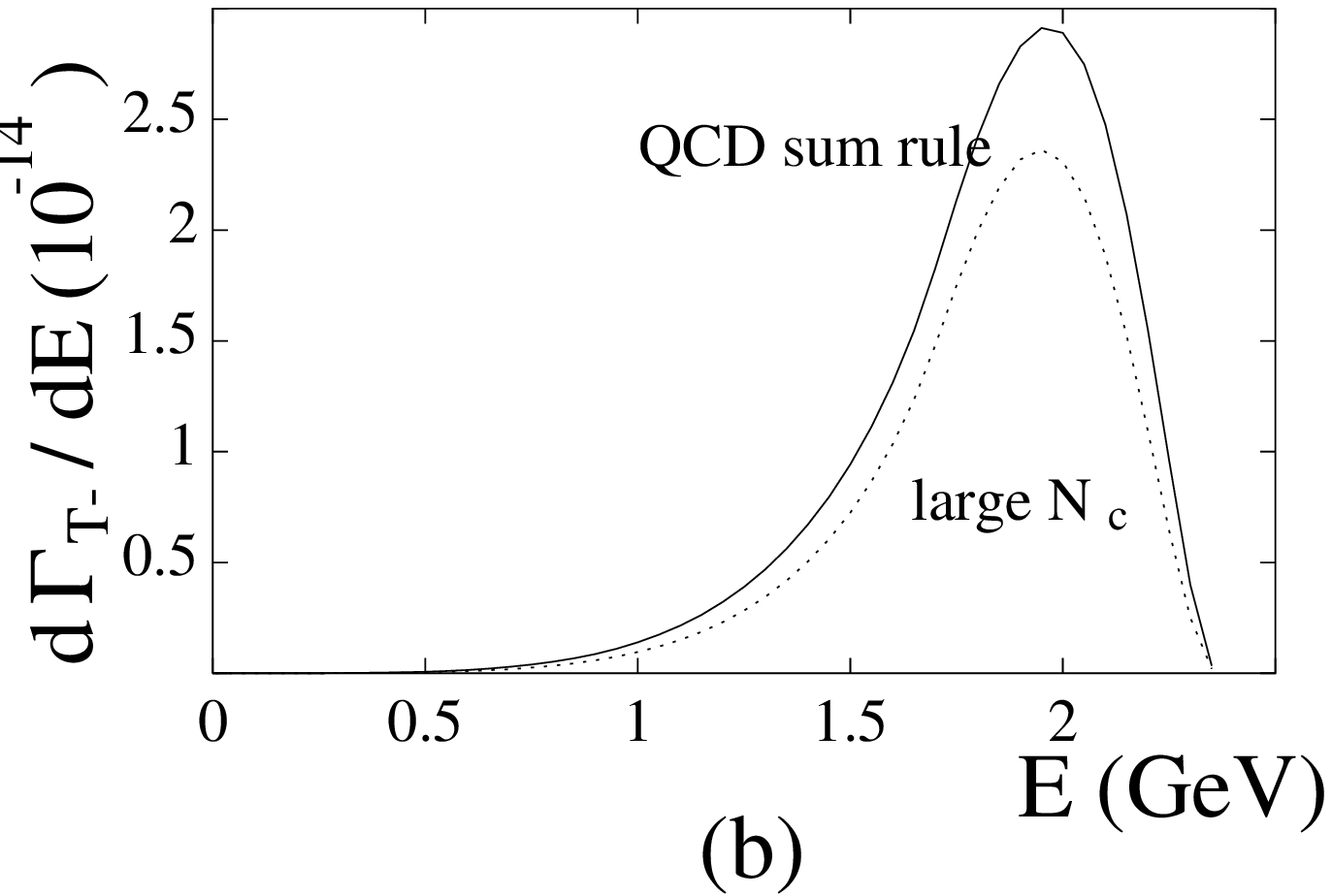, height=5cm}\\[5mm]
\epsfig{file=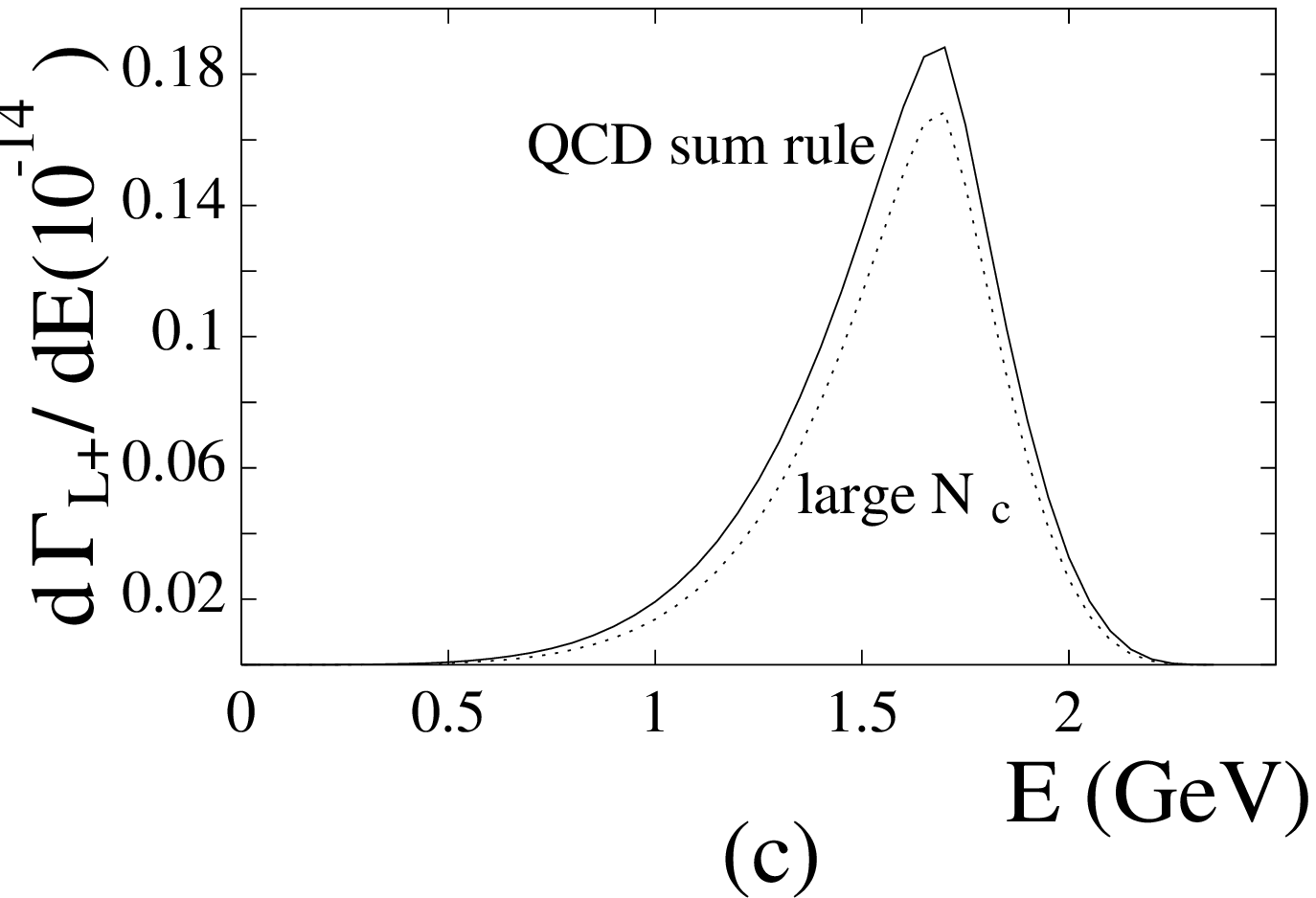, height=5cm}&
~~~\epsfig{file=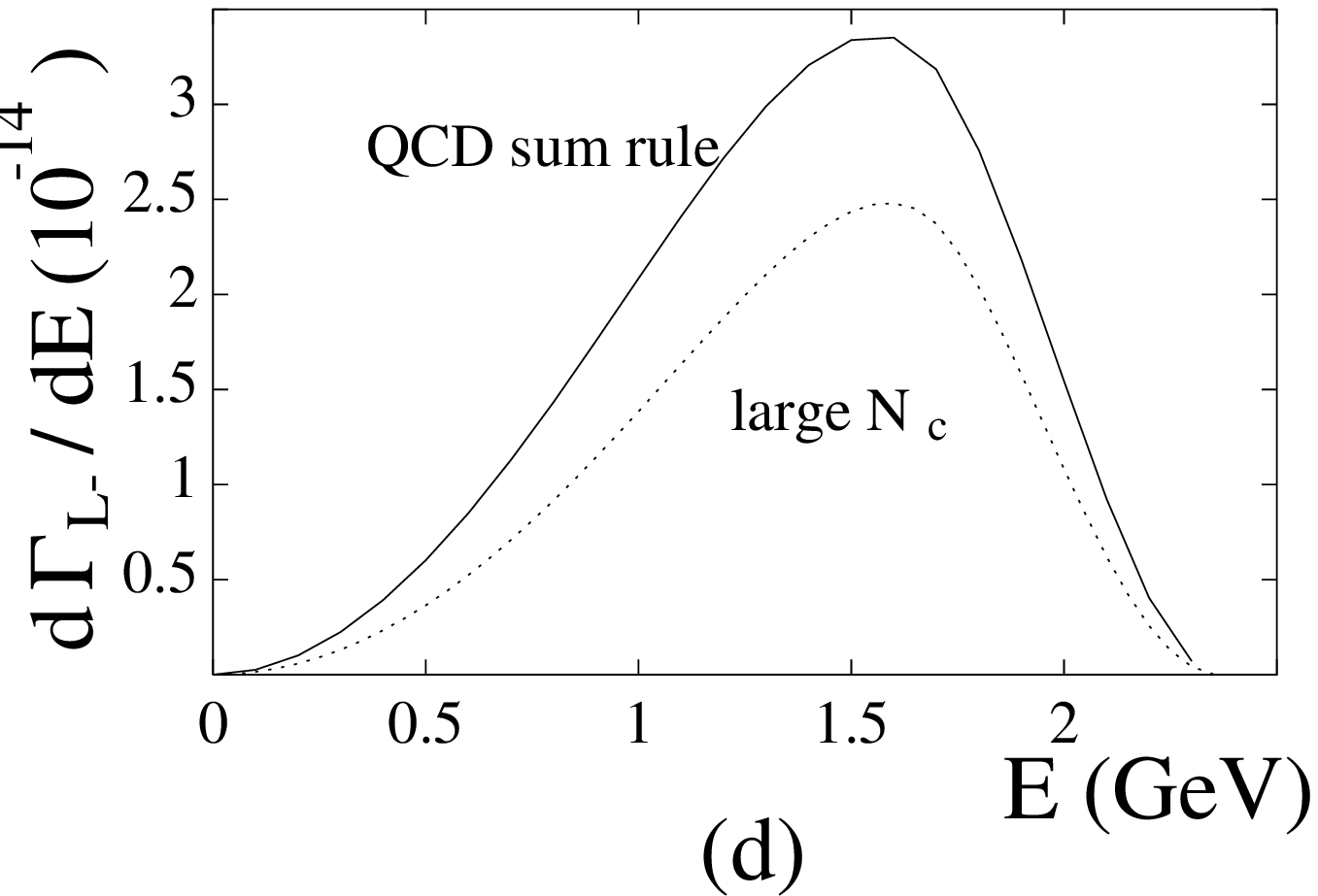, height=5cm}\\[5mm]
\epsfig{file=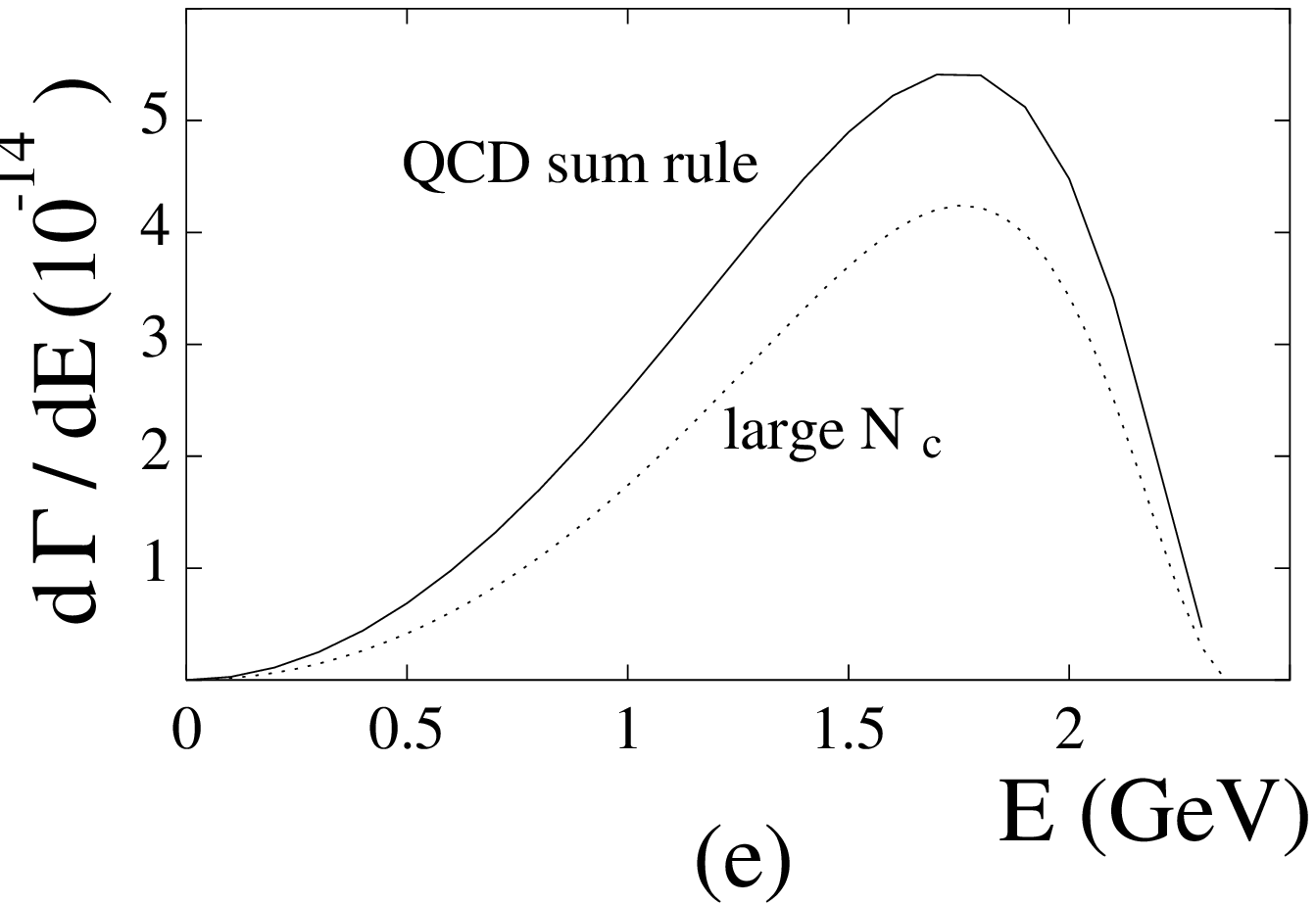, height=5cm}
\end{tabular}
\label{fig:ef}
\caption{}
\end{figure}
%==================================================================

%======================= TABLE1 =============================================
\begin{table}
\vskip 5mm
\begin{center}
\begin{tabular}{c|ccccc}
%\hline
  & $\Gamma_{T+}$  & $\Gamma_{T-}$ & $\Gamma_{L+}$ & $\Gamma_{L-}$ 
    & $\Gamma_{tot}$\\ \hline
 QCD sum rule & 0.432 & 1.86 & 0.103 & 3.77 & 6.17\\
 large $N_c$  & 0.343 & 1.45 & 0.087 & 2.63 & 4.51\\
 quark model \cite{Kroll} & 0.408 & 1.20 & 0.092 & 2.58 & 4.28
%\\ \hline
\end{tabular}
\end{center}
\label{table:rates}
\caption{}
\end{table}

%============================= TABLE 2 ======================================
\begin{table}
\vskip 3mm
\begin{center}
\begin{tabular}{c|cccccc}
  & $<\alpha>$ & $<\alpha^\prime>$ & $<\alpha^{\prime\prime}>$ & $<\gamma>$ 
  & $<\alpha_P>$ & $<\gamma_P>$ \\ \hline
  sum rule & -0.83 & -0.14 & -0.57 & 0.48 & 0.38 & -0.17\\ 
  large $N_c$ & -0.81 & -0.15 & -0.53 & 0.50 & 0.34 & -0.19\\ 
  quark model \cite{Kroll}
  & -0.77 &-0.11 & -0.54 & 0.55 & 0.40 & -0.16 
\end{tabular}
\end{center}
\label{table:asym}
\caption{}
\end{table}
%=============================================================================

%===================== TABLES(MULTICOLUMN) ============================
\begin{table}
\vskip 5mm
\begin{center}
\begin{tabular}{c|ccc|ccc}
 & \multicolumn{3}{c|}{$\Gamma(10^{10}~{\rm sec}^{-1})$}  
 & \multicolumn{3}{c}{$\alpha$} \\ 
\cline{1-7}
 Modes & sum & large & quark & sum & large & quark \\ 
 & rule & $N_c$ & model \cite{Cheng} 
   & rule & $N_c$ & model \cite{Cheng}\\ \hline
 $\Lambda_c\pi$ & $0.780$ & $0.406$ & $0.342$ & -1.00 & -1.00 & -0.996\\
 $\Lambda_c\rho$ & 1.08 & 0.583 & 0.489 & -0.886 & -0.885 & -0.876\\
 $\Lambda_c K$ & 0.057 & 0.030 & 0.025 & -1.00 & -1.00 & -0.997\\
 $\Lambda_cK^*$ & 0.055 & 0.030 & 0.025 & -0.853 & -0.853 & -0.842
\end{tabular}
\end{center}
\label{table:nonlep}
\caption{}
\end{table}
%==============================================================================


\begin{references}
%%%%%%%%%%%%%%%%%%%%%%%%%%%%%%%%%%%%%%%%%%%%%%%%%%%%%%%%%%%%%%%%%%%%%%%%%%%%%%%%
\bibitem{IsgurL}
N. Isgur and M.B. Wise, Phys. Lett. {\bf B 232} (1989) 113; {\bf 237} 
(1990) 527;\\
M.B. Voloshin and M.A. Shifmas, 
 Yad. Fiz. {\bf 45} (1987) 463, {\bf 47} (1988) 801;\\
E.V. Shuryak, Phys. Lett. {\bf B 93} (1980) 134. 

\bibitem{GeorgiL}
H. Georgi, Phys. Lett. {\bf B 240} (1990) 447;\\
A.F. Falk, H. Georgi, B.Grinstein and M.B. Wise, Nucl. Phys. {\bf B 343} 
(1990) 1.

\bibitem{WiseL}
N. Isgur and M.B. Wise, Nucl. Phys. {\bf B 348} (1991) 276;\\
H. Georgi, Nucl. Phys. {\bf B 348} (1991) 293;\\
T. Mannel, W. Roberts and Z. Ryzak, Nucl. Phys. {\bf B 355} (1991) 38.

\bibitem{LukeL}
M.E. Luke, Phys. Lett. {\bf B 252} (1990) 447.

\bibitem{GrinsteinL}
H. Georgi, B.Grinstein and M.B. Wise, Phys. Lett. {\bf B 252} (1990)
456;\\
Y.B. Dai, X.H. Guo and C.S. Huang, Nucl. Phys. {\bf B 412} (1994) 277.

\bibitem{Grozin}
A.G. Grozin and O.I. Yakovlev, Phys. Lett. {\bf B 291} (1992) 441.

\bibitem{Dai}
Y.B. Dai, C.S. Huang, M.Q. Huang and C. Liu,
Phys. Lett. {\bf B 387} (1996) 379;\\
Y.B. Dai, C.S. Huang, C. Liu and C.D. L\"u, 
Phys. Lett. {\bf B 371} (1996) 99.

\bibitem{ManoharL}
E. Jenkins, A.V. Manohar and M.B. Wise, 
Nucl. Phys. {\bf B 396} (1993) 38.

\bibitem{UKQCD}
UKQCD Collab., Nucl. Phys. Proc. Suppl. {\bf 53} (1997) 408; hep-lat/9709028.

\bibitem{Boyd}
C.G. Boyd, R.F. Lebed, Nucl. Phys. {\bf B 485} (1997) 275;\\ 
D. Chakraverty, T. De, and B. Dutta-Roy,
SIN-TNP/98-04, hep-ph/9802223.

\bibitem{IvanovL}
See for example, M.A. Ivanov, V.E. Lyubovitskij, J.G. K\"orner and 
P. Kroll, Phys. Rev. {\bf D 56} (1997) 348.

\bibitem{Shifman}
M.A. Shifman, A.I. Vainshtein and V.I. Zakharov, 
Nucl. Phys. {\bf B 147} (1979)385; {\bf B 147} (1979) 488.

\bibitem{Hooft}
G. 't Hooft, Nucl. Phys. {\bf B 72} (1974) 461;\\
E. Witten, Nucl. Phys. {\bf B 160} (1979) 57.

\bibitem{JenkinsL}
E. Jenkins, A.V. Manohar and M.B. Wise, Nucl. Phys. {\bf B 396} (1993) 27;\\
C.G. Callan and I. Klebanov, Nucl. Phys. {\bf B 262} (1985) 222;\\
M. Rho, D.O. Riska and N.N. Scoccola, Phys. Lett. {\bf B 251} (1990) 597;
Z. Phys. {\bf A 341} (1992) 343;\\
Y. Oh, D.-P. Min, M. Rho and N.N. Scoccola, 
Nucl. Phys. {\bf A 534} (1991) 493.

\bibitem{Chow}
C. Chow and M.B. Wise, Phys. Rev. {\bf D 50} (1994) 2135;\\
C. Liu, Phys. Lett. {\bf B 389} (1996) 347.
\bibitem{Koerner}
   J.G. K\"orner and M. Kr\"amer, 
     Phys. Lett. {\bf B 275} (1992) 495;\\
   P.Bialas, J.G. K\"orner, M. Kr\"amer and K. Zalewski,
     Z. Phys. {\bf C 57} (1993) 115.
\bibitem{Kroll}
   B. K\"onig, J.G. K\"orner, M. Kr\"amer and P. Kroll,
   Phys. Rev. {\bf D 56} (1997) 4282.
\bibitem{Bauer}
   M. Bauer, B. Stech and M. Wirbel,
   Z. Phys. {\bf C 34} (1987) 103.
\bibitem{Neubert2}
   For a recent review, see 
   M. Neubert and B. Stech, HD-THEP-97-23, hep-ph/9705292.
\bibitem{Cheng}
   H.Y. Cheng, Phys. Rev. {\bf D 56} (1997) 2799.
\bibitem{Rusetsky}
   M. A. Ivanov, J. G. K\"orner, V. E. Lyubovitskij, and A. G. Rusetsky,
   hep-ph/9709372; hep-ph/9710523.
\bibitem{Giri}
   A.K. Giri, L. Maharana and R. Mohanta,
   hep-ph/9710285.
\end{references}
\end{document}